\documentclass{mem}
\usepackage{natbib}\usepackage{txfonts}\usepackage{balance}
\usepackage{graphicx}
\usepackage[a4paper,breaklinks,dvipdfm]{hyperref}
\idline{00}{000}
\begin{document}
\def\kms{$\mathrm {km s}^{-1}$}

\title{Metal abundances in the high-redshift intergalactic medium}
\author{Michele \,Fumagalli\inst{1,2,3}}
 
\institute{
Carnegie Observatories, 813 Santa Barbara Street, Pasadena, CA 91101, USA. 
\email{mfumagalli@obs.carnegiescience.edu}
\and
Department of Astrophysics, Princeton University, Princeton, NJ 08544-1001, USA.
\and 
Hubble Fellow
}

\authorrunning{Fumagalli}

\titlerunning{IGM metal abundances at $z\gtrsim 2$}

\abstract{
Twenty years of high-resolution spectroscopy
at the $8-10$ m class telescopes have drastically expanded
our view of the gas-phase metallicity in the $z\gtrsim 2$ universe.
This contribution briefly summarizes how these studies reveal a widespread 
metal pollution in the intergalactic medium with a median 
abundance $\rm [C/H]\sim -3.5$ at $z \sim 3$ that is increasing by a 
factor of $\sim 2-3$ from $z\sim 4.3$ to $z\sim 2.4$. At the higher densities 
that are typical of galactic halos, observations uncover 
a metallicity spread of five orders of magnitude in Lyman limit systems, 
ranging from super-solar ($\rm [M/H] \sim +0.7$) to pristine ($\rm [M/H] \lesssim -4$)
gas clouds. Finally, the neutral damped Ly$\alpha$ systems are  
enriched to a median metallicity of $\rm [M/H] \sim -1.5$ that slowly declines 
with redshift up to $z\sim 4.5$, at which point it appears to more rapidly evolve
as one approaches the end of reionization.
}
\maketitle{}

\section{Introduction}

The study of metal lines associated to intervening absorption line systems (ALSs)
that are detected in the foreground of $z\gtrsim 2$ quasars offers a unique 
opportunity to trace the history of the metal enrichment of cosmic 
structures in a redshift-independent fashion, up to the first billion 
year from the Big Bang. Since the advent of high-resolution 
echelle and echellette spectrographs at the $8-10$ m class telescopes, 
observers have been able to map the chemical 
composition of the intergalactic medium (IGM) and of the denser gas within 
and around distant galaxies.

In this contribution, after a brief overview of the adopted techniques that may be useful to 
readers that are not familiar with measurement of gas metallicities in absorption
(Sect. \ref{sec:tech}), we summarize the current status of the study of the metal 
content in the Ly$\alpha$ forest 
(Sect. \ref{sec:lya}), in Lyman limit systems (LLSs, Sect. \ref{sec:lls}), 
and in damped Ly$\alpha$ systems (DLAs, Sect. \ref{sec:dla}). 
Given the large body of literature on the subject, this article cannot be considered a 
complete review since we focus primarily on observational results at $z\gtrsim 2$, 
only briefly referring to theoretical studies. Furthermore, we have decided to highlight 
empirical findings and their simple interpretation, without however examining in detail 
the physical processes that are responsible for the observed metal pollution of 
cosmic structures. Discussions on some of these aspects can be found in other 
chapters of these proceedings.

\section{Technique}\label{sec:tech}

\subsection{Counting atoms in the distant universe}

With absorption spectroscopy, it is straightforward to directly 
measure the abundances of several elements in different ionization 
states, provided that lines are not saturated.
For each transition, the transmitted flux of a background source is 
simply $I_{\rm tran} = I_{\rm back}\exp(-\tau(\nu))$, where the
optical depth is defined by $\tau(\nu) = \frac{\pi e^2}{m_e c} f N_{\rm kj} \phi(\nu,b)$.
Here, $f$ is the oscillator strength, $N_{\rm kj}$ is the ion column density, and
$\phi$ is the frequency-dependent line profile that also depends 
on the Doppler parameter $b$ \citep[e.g.][]{spi78}. Differently
from studies in emission where detection limits translate into 
distance-dependent sensitivity on physical quantities, one can notice from the above 
equations that the detection limits on column densities are conveniently redshift 
independent. This means that the metal content of intervening ALSs can be measured
with equal sensitivity in column density across over 10 Gyr of cosmic history. 

While column densities are directly observables, complications arise 
when metallicities are derived because of ionization corrections. 
Neutral gas is a dominant component only at the high hydrogen column 
densities observed in DLAs ($N_{\rm HI} \ge 10^{20.3}~\rm cm^{-2}$) for which 
metallicities are simply obtained by dividing the measured column densities 
in neutral or singly ionized ions by the neutral hydrogen column densities.
However, for the majority of ALSs at lower $N_{\rm HI}$, the observed ions trace only 
a (sometimes small) fraction of the underlying hydrogen and 
metal content. In these cases, ionization corrections  are required to translate the 
observed column densities into metallicities.
  
For the classes of ALSs that are discussed here, it can 
be generally assumed that gas is predominantly photo-ionized, as indicated for instance 
by the line ratios of a same element in adjacent ionization states 
(e.g. \ion{C}{II}, \ion{C}{III}, \ion{C}{IV}). Standard practice is also 
to assume that gas is in photoionization equilibrium \citep[but see][]{gna10,opp13}. 
Ionization corrections are then computed via one-dimensional radiative transfer 
calculations, with the {\tt cloudy} software package \citep{fer98} being the code of 
choice for many authors. 

To derive ionization corrections, the observed column densities $N_{kj}$ of 
elements $j$ in the $k-$th ionization states are compared to grids of 
column densities that are computed as a function of free parameters, such as the 
volume density $n_{\rm H}$, the spectrum of the incident radiation 
field $j(\nu)$, and the chemical composition $A_j$. For the set of parameters
that best describe the observations, the photoionization models 
yield the desired ionization corrections $X_{kj}$ such that $N_j = N_{kj} X_{kj}$.
Metallicities can then be inferred from tracer elements as
${\rm [M/H]} = \log (N_j/N_H) - \log(N_j/N_H)_\odot$.
 
Although in some cases volume densities can be constrained directly by 
observations (e.g. by transitions from fine-structure levels) or indirectly 
(e.g. requiring that gas is in hydrostatic equilibrium), the problem of establishing 
ionization corrections is often highly under-determined. Standard practice is therefore 
to assume that the gas clouds are illuminated by the extra-galactic ultraviolet 
background (EUVB) of a given spectral shape and intensity 
\citep[e.g.][]{fau09,haa12}. Further, it is often assumed that the gas has the same 
chemical abundance pattern of the solar neighborhood. Naturally, all these 
assumptions introduce systematic uncertainties in the analysis and metallicities 
become model dependent.

\subsection{From absorption systems to cosmic structures}

ALSs are broadly classified in three different classes, according to their observed 
neutral hydrogen column densities. Systems with $N_{\rm HI} \lesssim 10^{16}~\rm cm^{-2}$ 
constitute the Ly$\alpha$ forest \citep[e.g.][]{lyn71}, while systems with 
$10^{16}{~\rm cm^{-2}} \lesssim N_{\rm HI} < 10^{20.3}~\rm cm^{-2}$ are defined 
Lyman limit systems\footnote{Systems with $10^{16} 
{~\rm cm^{-2}} \lesssim N_{\rm HI} \lesssim 10^{17.2}~\rm cm^{-2}$ are typically classified 
as partial Lyman limit systems (pLLSs), but we do not make this distinction here.} 
\citep[LLSs; e.g.][]{tyt82} or, at the higher column densities ($N_{\rm HI} \gtrsim 10^{19}~\rm cm^{-2}$), 
super Lyman limit systems \citep[SLLSs; e.g.][]{per02}. Finally, 
systems with $N_{\rm HI} \ge 10^{20.3}~\rm cm^{-2}$ are defined damped Ly$\alpha$ 
systems \citep[DLAs; e.g.][]{wol05}. 

Unfortunately, through absorption spectroscopy alone, observers cannot directly 
connect the individual gas clouds detected against background sources to the 
physical structures that host them. However, it is possible to establish a 
general correspondence between these three classes of ALSs and astrophysical structures 
by assuming that gas clouds are in local hydrostatic equilibrium \citep[e.g.][]{sch01}. The validity 
of this assumption, that is also supported by hydrodynamic simulations 
\citep[e.g.][]{cen94,rah13}, can be simply understood in the following way. If gas clouds 
were out of equilibrium, then they would expand until equilibrium is restored with the 
ambient medium, or collapse and fragment into smaller structures of sizes comparable to the 
Jeans length $L_J$. It follows that volume densities and column 
densities are related by $N_{\rm H} = L_J n_{\rm H}$, where $N_{\rm H} = N_{\rm HI} X_{\rm HI}$ as 
set by radiative transfer processes. Given these two simple equations, from a 
theoretical point of view, 
the Ly$\alpha$ forest can be identified with gas densities $n_{\rm H} \lesssim 10^{-3}~\rm cm^{-3}$ 
that are typical of the cosmic web and voids \citep[e.g.][]{cen94,her96}, 
LLSs with gas densities $10^{-3}{~\rm cm^{-3}} \lesssim n_{\rm H} \lesssim 10^{-1.5}~\rm cm^{-3}$
that are comparable to or above virial densities \citep[e.g.][]{koh07,fum11}, 
and DLAs with gas densities $n_{\rm H} \gtrsim 10^{-1.5}~\rm cm^{-3}$ that are commonly  
found in galactic disks or neutral gas clumps \citep[e.g.][]{nag04,pon08}. 
For reference, the mean cosmic density at $z\sim 3$ is $\bar n_{\rm H} \sim 10^{-5}~\rm cm^{-3}$. 
For this reason, ALSs are powerful probes of the cosmic metal content and its evolution.

\section{Metals in the Ly$\alpha$ forest}\label{sec:lya}

\subsection{The metallicity of the IGM}

The metal content of the IGM can be inferred from the analysis of the Ly$\alpha$ forest that is 
comprised by gas clumps with $N_{\rm HI} \lesssim 10^{16}~\rm cm^{-2}$. Three independent 
techniques have been adopted in the past to measure the IGM metallicity:
the pixel optical depth method \citep[POD; e.g.][]{cow98,ell00,sch03}, 
the study of individual Ly$\alpha$ forest lines \citep[e.g.][]{dav98,car02,sim04,sim11}, and the 
analysis of composite quasar spectra \citep[e.g.][]{lu998,pie10,pie13}.  

The POD method relies on a statistical analysis of multiple quasar sightlines 
\citep[see e.g.][]{sch03}. At first, the optical depths of the 
\ion{C}{IV} ion $\tau_{\rm CIV}$ and of the associated neutral 
hydrogen $\tau_{\rm HI}$ are measured at each spectral pixel 
inside the wavelength windows where \ion{C}{IV} absorption lies. 
After applying corrections for noise and 
contaminants, temperatures and densities are assigned to each value of $\tau_{\rm HI}$ 
using cosmological hydrodynamic simulations. It should be noted that this type of 
simulations are fairly robust \citep{the98,reg07} and less dependent on the often 
unknown sub-grid physics that is adopted in simulations of galaxy formation. 
Therefore, the use of simulations is not a source of large and uncontrolled 
systematic uncertainties. 

The last step in the derivation of the Ly$\alpha$ forest metallicity is to apply  
ionization corrections $X_{\rm CIV}$ and $X_{\rm HI}$ to infer the total carbon abundance: 
\begin{equation}
{\rm [C/H]} = \log \left(\frac{\tau_{\rm CIV}\lambda_{\rm CIV} f_{\rm CIV} X_{\rm CIV}}
{\tau_{\rm HI}\lambda_{\rm HI} f_{\rm HI} X_{\rm HI}}\right) - (\rm C/H)_\odot\:.
\end{equation}
This last step is the most prone to non-negligible systematic uncertainties \citep[see][]{sim11}, also  
because \ion{C}{IV} is a sub-dominant ion at most densities, 
being $\lesssim 10\%$ of all the carbon present inside the IGM at $z\sim 2.5$. 
Furthermore, ionization corrections systematically vary as a function of the 
assumed EUVB spectral shape, which is uncertain at the energies that are responsible 
for the photoionization of doubly and triply ionized carbon. Therefore, for the 
same observed \ion{C}{IV} distribution, an EUVB with harder spectral shape 
\citep[e.g.][]{haa01} implies higher carbon abundances at low densities 
compared to what one would infer using a softer spectrum \citep[e.g.][]{fau09}. 

Following this procedure, \citet{sch03} measured the distribution of carbon within the 
Ly$\alpha$ forest as a function of overdensity $\delta=n_{\rm H}/ \bar n_{\rm H}$ 
between $z\sim 2-4$, with most of the data lying at $z\sim 3$. For the assumed 
\citet{haa01} EUVB and solar carbon abundance ${\rm (C/H)_\odot = -3.45}$ \citep{and89},
the median IGM metallicity is
${\rm [C/H]} = -3.47^{+0.07}_{-0.06} + 0.08^{+0.09}_{-0.10}(z-3)+
0.65^{+0.10}_{-0.14}(\log \delta -0.5)$ with a log-normal scatter 
${\rm \sigma([C/H])} = 0.76^{+0.05}_{-0.08} + 0.02^{+0.08}_{-0.12}(z-3)
-0.23^{+0.09}_{-0.07}(\log \delta -0.5)$. 

The second method to establish the metal content of the IGM relies on the 
identification of individual metal lines that are associated to distinct ``clouds'' 
in the Ly$\alpha$ forest. By fitting the hydrogen Lyman 
series and the associated strong metal lines such as \ion{C}{IV} and \ion{O}{VI}, 
this analysis provides a more direct constraint on the IGM metal 
enrichment than the POD method. However, these types of study are limited to overdensities 
$\delta \gtrsim 2$ because of the minimum hydrogen column densities that can be 
detected even in the highest resolution and signal-to-noise spectra. 
Similarly, detection limits hamper the direct measurement of very low metallicities 
(e.g. $\rm [O/H] < -3$) in individual ALSs, although the underlying metal distribution 
of the IGM can be inferred using survival analysis.

With direct measurements, \citet{sim04} obtained a median IGM metallicity 
of $\rm [O/H]=[C/H]=-2.82$ with $\sim 0.75$ dex scatter at $z\sim 2.5$,  
after applying ionization corrections using an unpublished version 
of the Haardt \& Madau EUVB (HM1.80) and assuming that the observed 
gas clouds are in local hydrostatic equilibrium. It should be noted that the agreement 
between the median oxygen and carbon abundances does not imply 
that $\rm [C/O]=0$ in the IGM, as this ratio depends on the 
assumed shape of the EUVB, with softer spectra yielding lower $\rm [C/O]$ ratios. 
Compared to the results of \citet{sch03}, the analysis of \citet{sim04} favors a 
weaker density dependence for the IGM metallicity, although the 
distributions of carbon that are inferred either by direct detection of
individual lines or by the POD method lie in satisfactory agreement once systematic 
variations due to the different choices of EUVB are taken into account. 

All combined, these studies reveal a widespread metal pollution of the IGM at $z\sim 2.5-3$, 
with non-zero metallicity extending to under-dense regions with $\log \delta = -0.5 - 0.0$. 
However, about $30\%$ of the Ly$\alpha$ forest lines detected above $\delta \gtrsim 1.6$ 
exhibit a metallicity $\rm [C,O/H] \lesssim -3.5$, which excludes the presence of a metallicity 
floor in the IGM. By mass, $40-60\%$ of the universe is enriched above 
$\rm [O/H] \gtrsim -3$ while, by volume, only $20\%$ of the overdense universe 
has a metallicity $\rm [C/H] \gtrsim -3$. Integrated over the density distribution 
predicted by numerical simulations between $\log \delta = -0.5 - 2.0$, the carbon 
density of the $z\sim 3$ universe is $\Omega_{\rm C} \sim 2.3 \times 10^{-7}$, for an 
assumed baryon density $\Omega_{\rm b} =0.045$ and a mass-weighted metallicity 
$\rm [C/H] = -2.80 \pm 0.13$. These estimates bear significant uncertainties related 
to ionization corrections, as well as extrapolations outside the range of 
densities directly probed by observations.  

More recently, \citet{sim11} has extended the analysis of the chemical properties of the
overdense ($\delta > 1.6$) Ly$\alpha$ forest to $z\gtrsim 4.0$, finding a median 
$\rm [C/H]=-3.55$ with log-normal scatter $\sigma \sim 0.8$ dex for the assumed 
\citet{gre98} solar abundances. While ionization corrections still constitute 
a source of systematic uncertainty, \ion{C}{IV} traces $\sim 50-60\%$ 
of all the carbon present in the IGM at $z\gtrsim 4$ for the 
mean \ion{H}{I} column density that is detectable in the Ly$\alpha$ forest. 
And in fact, consistent median abundances are found for either the 
\citet{haa01} or the \citet{fau09} EUVB. Compared to carbon and oxygen 
abundances at similar overdensities in the lower redshift universe, the 
median IGM metallicity at $z\gtrsim 4-4.5$ is a factor of $\sim 2-3$ lower
than what observed at $z\sim 2.5$. As this evolution is only 
weakly dependent on the assumed EUVB,  this analysis implies that half of the 
metals seen in the IGM at $z \sim 2.5$ are ejected from galaxies  
between $z \sim 2.5-4.3$. This value is indicative of a substantial pre-enrichment of 
the IGM, and it implies that the observed metal content of the Ly$\alpha$ forest at a given redshift
is not entirely attributable to coeval star-formation episodes.

\subsection{The structure of metals within the IGM}

The study of the small-scale metal distribution in the IGM offers interesting clues to the 
processes that are responsible for the observed metal enrichment.
A direct way of establishing the small-scale spatial distribution of metals in the 
IGM is to analyze the variation in the metal line profiles along 
closely-spaced quasar sightlines with projected separations between 
$\sim 10$ pc and a few kpc. By examining spectra of
multiple images of strongly lensed quasars, \citet{rau99} and \citet{rau01} concluded 
that low-ionization lines at $z\sim 2-3.5$ (e.g. \ion{C}{II}, \ion{Si}{II}) arise 
from very compact clouds with sizes as small as $\sim 30~\rm pc$. Conversely, metal lines 
of triply-ionized carbon or silicon appear featureless on scales of few hundred 
parsecs, and they retain a coherence in velocity up to a few kpc. Similar sizes
are also found in photoionization modeling of metal lines detected in individual 
quasar spectra \citep[e.g.][]{sch07,pie13}.

Compared to the Jeans scale of the high-redshift Ly$\alpha$ forest 
($\sim 100$ kpc), the existence of enriched gas clumps with sizes of few hundred parsecs
implies that metals are not fully mixed with the surrounding hydrogen distribution, 
at least on timescales that are comparable to the lifetime of the enriched clouds. 
Given that the implicit assumption to any metallicity determination is that the observed metal
and hydrogen column densities arise from the same phase, it follows that the inferred metallicities 
in the Ly$\alpha$ forest are often smoothed on the larger spatial scales of hydrogen.  

As for the origin of the observed enrichment, the compact and 
metal-rich clouds ($\rm [C/H]>-1$) that are uncovered by the analysis of \citet{sch07} 
outnumber galaxies by several order of magnitudes. Also, these clouds are likely to be 
short-lived ($\sim 10^7$ yr), either because they lack pressure support from an ambient 
medium or because of the hydrodynamic instabilities.  It is therefore plausible that a 
significant fraction of the metals seen in the IGM has been carried from galaxies in the form of 
compact and enriched clumps that subsequently mix with the surrounding hydrogen. 
Multi-phase galactic winds \citep[e.g.][]{cre13} or stripped gas clouds 
\citep[e.g.][]{rau13} are two possible channels for the IGM pollution.

\section{Metals in Lyman limit systems}\label{sec:lls}

The study of the metal content of LLSs ($N_{\rm HI} \sim 10^{16}-10^{20.3}~\rm cm^{-2}$)
provides valuable insights into the chemical properties of gas at the densities that are comparable to or 
higher than the virial densities. A detailed
analysis of the metallicities of LLSs is therefore a powerful way to probe the flows of 
metals in galaxy halos, at the interface between galactic disks and the IGM 
\citep[e.g.][]{fum11,fau11,van12,rud12,fum13}. 
However, the chemical abundances of LLSs have been analyzed in sufficiently large samples 
only at $N_{\rm HI} \gtrsim 10^{19}~\rm cm^{-2}$, the column densities  that are typical for SLLSs 
\citep[e.g.][]{per03,kul07}. Between $N_{\rm HI} \sim 10^{16}-10^{19}~\rm cm^{-2}$, where the bulk of the 
LLS population resides, only the metal properties of few individual systems or small samples 
have been reported so far \citep[e.g.][]{ste90,pro99,fum11b,fum13a}. Such a lack of
comprehensive studies (which are however underway) is imputable to the fact that 
high signal-to-noise and resolution data are required to measure $N_{\rm HI}$ in systems which  
have most of the Lyman series lines saturated.

\citet{fum13a} have recently obtained a composite spectrum of a small statistical sample of LLSs
 at $z\sim 2.6-3.0$ \citep[see also][]{prok10}, which is characterized by 
prominent absorption lines of triply-ionized carbon and silicon 
with equivalent widths in excess to 0.5\AA. Higher-ionization lines, e.g. \ion{C}{IV} and 
\ion{Si}{IV}, are weaker with equivalent widths of $\sim 0.1-0.2$ \AA, while lower-ionization lines 
(e.g. \ion{O}{I}, \ion{Si}{II}, \ion{C}{II}) are mostly undetected to limits of $\sim 0.05-0.1$ \AA\ 
 below $N_{\rm HI} \lesssim 10^{19}~\rm cm^{-2}$. All combined, these properties 
are indicative of an highly ionized\footnote{The ionization parameter $U$ is defined by the ratio of 
hydrogen ionizing photons to the total hydrogen density.} ($\log U \gtrsim -3$) and likely metal poor 
gas phase 
($\rm [M/H]\lesssim -1.5$), as confirmed by photoionization modeling under the assumption that LLSs are 
illuminated by the \citet{haa12} EUVB and have solar abundance ratios \citep{asp09}. Metallicities between 
$\rm -3 \lesssim [M/H] \lesssim -1.5$ are also commonly found in the analysis of individual LLSs \citep{ste90}.

Although we currently lack a well sampled metallicity distribution, a striking peculiarity of this 
class of absorbers is the observed range of five orders of magnitude in the LLS metal content. 
At one extreme of the metallicity distribution, \citet{pro06} found a SLLS with metallicity 
$\rm [M/H]\sim +0.7$, a value that only moderately depends on ionization corrections 
(with amplitude $\sim 0.3$ dex) given the observed column density of $N_{\rm HI} = 10^{19}~\rm cm^{-2}$. 
At the other extreme, \citet{fum11b} reported the discovery of two gas clouds without any associated 
heavy metals. In the absence of metal lines, the ionization state of the gas is unconstrained but, 
under the conservative assumption that these gas clouds have $\log U \gtrsim -3$, the 
metallicities of these two pristine systems are $\rm [M/H] \lesssim -4$.  Notably, one of the two 
clouds also has a deuterium abundance of primordial composition. This clump is therefore a relic of 
the gas that formed during the Big Bang nucleosynthesis and remained uncontaminated for 
two billion years. 

Besides the large metal variation from system to system, LLSs 
also exhibit heterogeneous chemical properties within individual absorbers. Few examples of 
LLSs with multiple line components that are separated by only a few hundreds $\rm km~s^{-1}$ 
but have metallicity differences up to 2 dex are accumulating in the literature 
\citep[e.g.][]{dod01,pro10,cri13}. Such a diversity of chemical composition both in individual 
systems and within the population suggests that the denser ambient gas in 
proximity to galaxies, from which LLSs likely arise, 
is composed by multiple co-existing but distinct gas phases, such as 
chemically-pristine inflows and metal-enriched outflows or galactic fountains.

\section{Metals in damped Ly$\alpha$ systems}\label{sec:dla}

At the highest neutral hydrogen column densities of $N_{\rm HI} \gtrsim 10^{20.3}~\rm cm^{-2}$,
quantum mechanic effects imprint a characteristic shape to the hydrogen Ly$\alpha$ profile that
can be used to precisely measure $N_{\rm HI}$ in individual systems. Furthermore, at these
densities, gas is almost fully neutral, thus neutral or singly ionized ions are the dominant 
species that trace the bulk of the gas. For these reasons, measurements of metallicities in 
DLAs are generally reliable, being unaffected by the uncertainties that arise from 
ionization corrections. Many studies 
have investigated in detail the metal content of DLAs, their 
redshift evolution, 
and their abundance ratios \citep[e.g.][]{pet02,pro03,raf12,nee13,jor13}.   

These studies, and in particular recent work by \citet{raf12}, have revealed a metallicity distribution 
that is well described by a Gaussian with mean $\rm [M/H] =-1.51$ and dispersion $\sigma = 0.57$. This 
distribution lacks extended tails either towards super-solar or pristine values, as also 
confirmed by dedicated searches for very low-metallicity DLAs \citep[e.g.][]{pen10} that consistently
found gas enriched at or above a floor of $\rm [M/H] \sim -3$. Further, the hydrogen-weighted mean 
metallicity of DLAs $\langle Z \rangle$, a quantity which describes the cosmic metal content of 
neutral hydrogen, exhibits a statistically-significant redshift evolution that \citet{raf12}
modeled as $\langle Z \rangle = (-0.22 \pm 0.03) z - ( 0.65 \pm 0.09) $ between $z\sim 0-4.5$,
although other parametrizations can be found in the literature \citep{jor13}.
Further, the study of the handful of DLAs known beyond $z\sim 4.5$ reveals that the 
cosmic metal content of neutral gas evolves much more rapidly in the first Gyr of cosmic 
history \citep{raf13}. This trend suggests a rapid chemical enrichment of 
the first cosmic structures or that a second population of more metal-poor DLAs becomes 
progressively important towards the epoch of reionization. 

Since ionization corrections are not needed, the intrinsic abundance patterns in DLAs
can be reliably measured \citep[e.g.][]{pro03b}. For the subset of systems in which both iron 
and $\alpha-$elements are measured, DLAs are found to be $\alpha-$enhanced with a median 
$\rm [\alpha/Fe] \sim +0.3$ \citep{raf12}. Furthermore, the $\rm [M/H]$ and [$\alpha$/Fe] 
distributions in $z>2$ DLAs with $\rm [M/H]\lesssim -1$ agree with those observed for 
Galactic halo stars. Similar abundance patterns between DLAs and halo stars are also 
found in the most metal-poor DLAs at $\rm [M/H] \lesssim -2$ 
\citep[e.g.][]{coo11}, hinting that there exists a connection between DLA gas and the sites where 
Galactic halo stars form. The study of abundance patterns of DLAs also provides important clues 
about the stellar populations that are responsible for the observed metal enrichment. 
For instance, a recent analysis by \citet{bec12} in candidate DLAs at $z>5$ revealed a remarkable 
similarity in the abundance ratios of DLAs from $z\sim 2$ to beyond $z\sim 6$. Given the age of 
the universe at $z\sim 6$, these observations indicate that supernovae type-II from population-II 
stars and prompt type-I supernovae are the major contributors to the observed metal 
enrichment in DLAs. 

\section{Summary}

Over two decades of absorption line studies have dramatically improved our view of the cosmic enrichment 
in the distant universe. Although a complete picture of the cosmic chemical evolution is yet 
to be drawn, this brief review outlines a possible sketch towards this goal. Galaxies are rapidly 
and continuously polluted by the yields of population II stars starting from the first few hundred million 
years of cosmic histories, and the relics of this enrichment are possibly seen today in the older stellar 
population within our Galaxy. Metals are also dispersed outside galactic disks, either through 
galactic winds or via more violent processes (e.g. tidal or ram-pressure stripping), in a 
heterogeneous, poorly mixed, and possibly short-lived gas phase. And this pollution is ultimately 
responsible for the widespread metal content that is visible in the overdense regions of the IGM.

\begin{acknowledgements}
The author thanks M.~Rafelski, X. Prochaska, and R. Simcoe for helpful discussions and comments on 
this manuscript and acknowledges support by NASA through the Hubble Fellowship grant 
HF-51305.01-A.
\end{acknowledgements}

\bibliographystyle{aa}

\end{document}